# On bifurcation behavior of hard magnetic soft cantilevers

Amir Mehdi Dehrouyeh-Semnani[1]

School of Mechanical Engineering, College of Engineering, University of Tehran, Tehran, Iran

Abstract
 Hard magnetic materials belong to a novel class of soft active materials with the capability of quick, large, and complex deformation via applying an external actuation. They have an extensive range of potential applications in soft robots, biomedical devices, and stretchable electronics, etc. Recently, investigation on the experimental and theoretical nonlinear mechanics of hard magnetic soft cantilevers has received notable considerations. In the available analyses, the most attention was paid to the non-bifurcation-type nonlinear mechanics of the system, and bifurcation was considered for a specific case. In the current study, the general trend of bifurcation-type nonlinear mechanics of hard magnetic soft cantilevers is introduced and the effective parameters on the occurrence of bifurcation are identified. Additionally, new trajectories due to the bifurcation and the corresponding workspace are presented. Eventually, a comprehensive comparative study between the new trajectories and workspace with those reported in the prior studies is carried out.

## 1. Introduction

   Soft active materials have attracted a great number of research interests from both scientific and engineering groups owing to their applications in the fields of soft robotics, stretchable electronic devices, wearable devices, biomedical devices. There exist several kinds of soft active materials that can deform under different external fields, such as electric, thermal, light, and magnetic fields. Hard magnetic soft active materials [1-5], as a new class of soft active materials, consist of a soft polymer matrix embedded with hard magnetic particles, or particles of high-coercivity ferromagnetic e.g. neodymium–iron–boron. They have the ability of fast, large, and complex transforming actuating, untethered control, and excellent programmability [6]. Therefore, the hard magnetic materials have been lately employed to construct soft active structures. Zhao et al. [7] investigated experimentally the extremely large deformation of the hard magnetic soft cantilevers with different aspect ratios under uniform perpendicular and antiparallel magnetic fields with different strengths. Additionally, an eight-node continuum brick element in the finite-element software Abaqus/Standard by writing a user-element (UEL) was developed to numerically analyze the nonlinear mechanics of the system. The results of the finite element model were in very good agreement with the experimental data. Wang et al. [8] developed an exact geometric nonlinear

---

[1] Email addresses: a.m.dehrouye@ut.ac.ir, a.m.dehrouye@gmail.com (A.M. Dehrouyeh-Semnani)

theory in conjunction with an explicit analytical solution for predicting the large deformation of hard magnetic soft beams under an actuating magnetic field. The comparative studies with the results of finite element simulations and experimental data validated the mathematical model and solution procedure. In the framework of the new model and solution strategy, the trajectories of the free-end location of the beam and the corresponding workspace as an end effector of soft robots were theoretically obtained and analyzed. Wu et al. [9] exhibited the applications of hard magnetic soft materials in basic one- and two-dimensional active structures with asymmetric configuration-shifting to biomimetic crawling robots, swimming robots, and two-dimensional metamaterials with tunable characteristics. Chen and Wang [10] studied analytically the nonlinear response of hard magnetic soft beams by considering the exact geometric nonlinearly and hyperplastic model for the constituent material of beam. Chen et al. [11] designed theoretically different deformed configurations of hard magnetic soft beams under an actuating magnetic field including, m-, s-, o-, $\alpha$-, and Ω- shapes, by considering piecewise inconstant residual magnetic flux density in the beam length direction and the exact geometric nonlinearity. Chen et al. [12] presented a theoretical model with the consideration of the exact geometric nonlinearity for the hard magnetic soft beams with continuously variable residual magnetic flux density and elastic modulus along the beam length direction. The linear buckling instability, nonlinear post-buckling, and nonlinear bending of system under linear and exponential variations of particle volume fraction was theoretically investigated via the Galerkin technique.

In the existing studies related to the nonlinear mechanics of hard magnetic soft cantilevers, the bifurcation was predicted only for the system subjected to an actuating antiparallel magnetic field. The current investigation aims to explore bifurcation-type responses of the hard magnetic soft cantilevers under an actuating magnetic field with an arbitrary angle. In Section 2, the exact geometric nonlinear formulation of a hard magnetic soft cantilever under an external magnetic field is presented and then a numerical solution based on the nonlinear shooting method is proposed. In Section 3, the experimental and theoretical results of previous studies are utilized to verify the present solution procedure. In Section 4, bifurcation-type mechanics of the system is studied and then new trajectories and workspace for the hard magnetic soft cantilevers are introduced and analyzed. The paper concludes with Section 5, where the work is summarized and the final remarks are presented.

## 2. Mathematical model and solution procedure

Figure1 illustrates the schematic representation of undeformed and deformed configurations of a hard magnetic soft beam exposed to an applied magnetic field. The elastic straight beam has a length of $L$, a cross section area of $A$, a second moment area of $I$, a Young's modulus of $E$, a shear modulus of $G$. The residual magnetic flux density of the beam in the undeformed and deformed configurations are denoted $B_0^r$ and $B^r$, respectively. It is supposed that $B_0^r$ is aligned with the $x$-direction. The magnetic flux density of the applied magnetic field is denoted $B^a$ and its angle with respect to the x-direction is denoted $\alpha$.

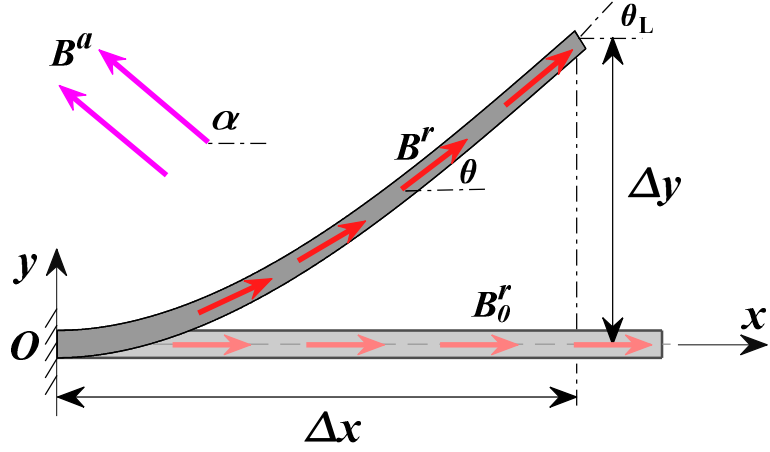

Figure1: Schematic representation of undeformed and deformed configurations of a hard magnetic soft cantilever under an external magnetic field.

The Mathematical model of hard magnetic soft beams is obtained under the following assumptions: (i) only the planer motion of the beam is considered; (ii) the elasto-dynamical model under the isothermal conditions is established; however, the quasi-static model is considered for the analysis. (iii) the beam is slender, the shear deformation and rotary inertia are ignorable, and therefore the nonlinear Euler-Bernoulli beam theory is applicable to establish exact formulation; (iv) the strain in the beam is taken small, even though large deformations are expected; (v) the constituent material of beam is incompressible and it behaves like a linear isotropic material.

In the framework of the Euler-Bernoulli beam theory, the only strain component developed in the longitudinal direction of the beam ($\varepsilon_{xx}$) in terms of the centerline strain ($\varepsilon$) and the centerline rotation angle ($\theta$) can be derived as

$$\varepsilon_{xx} = \varepsilon - y \frac{\partial \theta}{\partial x} \tag{1}$$

According to such longitudinal strain, the longitudinal stress of beam ($\sigma_{xx}$) can be expressed as

$$\sigma_{xx} = E \varepsilon_{xx} \tag{2}$$

The elastic energy of the beam ($U_s$) due to the longitudinal and bending deformations can be obtained as

$$U_s = \frac{1}{2} \int_V \sigma_{xx} \varepsilon_{xx} = \frac{1}{2} \left( \int_0^L EA \varepsilon^2 dx + \int_0^L EI \left( \frac{\partial \theta}{\partial x} \right)^2 dx \right) \tag{3}$$

The magnetic potential energy of the beam per unit volume of the deformed configuration can be expressed as

$$\phi_m = -\frac{1}{\mu_0} F B_0^r . B^a \tag{4}$$

in which $\mu_0$ stands for the vacuum permeability. Besides, $F$ represents the deformation gradient tensor. In the case of the planner displacement, $F$ takes the following form

$$F = (1+\varepsilon)\begin{bmatrix} \cos(\theta) & 0 \\ \sin(\theta) & 0 \end{bmatrix} \tag{5}$$

Additionally, $B^a$ and $B_0^r$ are formulated as follows

$$B_0^r = \begin{bmatrix} |B_0^r|, & 0 \end{bmatrix}^T, \quad B^a = |B^a|\begin{bmatrix} \cos(\alpha), & \sin(\alpha) \end{bmatrix}^T \tag{6a-b}$$

It should be pointed out that the residual magnetic flux density of the beam in the deformed configuration can be obtained by $B^r = FB_0^r$.

Inserting Eqs. (5) and (6a-b) into Eq. (4) gives

$$\phi_m = -(1+\varepsilon)\frac{|B_0^r||B^a|}{\mu_0}\cos(\alpha - \theta) \tag{7}$$

Considering the first part of the assumption (v), the magnetic potential energy of the beam can be expressed as

$$U_m = \int_V \phi_m dV = -\int_0^L \frac{A}{\mu_0}|B_0^r||B^a|(1+\varepsilon)\cos(\alpha - \theta)dx \tag{8}$$

In light of the deformed configuration of the beam illustrated in Figure1, the longitudinal displacement ($u$) and the lateral displacement ($v$) of the beam centerline can be expressed as

$$\frac{\partial u}{\partial x} = (1+\varepsilon)\cos(\theta) - 1, \quad \frac{\partial v}{\partial x} = (1+\varepsilon)\sin(\theta) \tag{9a-b}$$

Integrating once from Eq. (9a-b) and considering zero value for the centerline strain at $x=0$, the longitudinal and lateral displacements in terms of $\varepsilon$ and $\theta$ can be rewritten as

$$u = \int_0^x ((1+\varepsilon)\cos(\theta) - 1)dx, \quad v = \int_0^x (1+\varepsilon)\sin(\theta)\, dx \tag{10a-b}$$

Considering Eq. (10a-b) and assuming $\partial \varepsilon / \partial t$ is negligible as well as supposing $(1+\varepsilon)\partial \theta / \partial t \approx \partial \theta / \partial t$; the kinetic energy of beam ($T$) can be achieved by

$$T = \frac{1}{2}\int_0^L \rho A \left(\int_0^x \frac{\partial \theta}{\partial t}\cos(\theta)dx\right)^2 dx + \frac{1}{2}\int_0^L \rho A \left(\int_0^x \frac{\partial \theta}{\partial t}\sin(\theta)dx\right)^2 dx + \frac{1}{2}\int_0^L \rho I \left(\frac{\partial \theta}{\partial t}\right)^2 dx \tag{11}$$

Taking into account the elastic energy of the beam ($U_s$), the magnetic potential of the beam ($U_m$), and the kinetic energy of the beam ($T$) and implementing the Hamilton's principle ($\delta \int_{t_1}^{t_2}(T - U_s + U_m)dt = 0$), one can acquire the following governing equations

$$\delta\varepsilon: E\varepsilon - \frac{1}{\mu_0}\left|B_0^r\right|\left|B^a\right|\cos(\alpha-\theta) = 0$$

$$\delta\theta: \rho I \frac{\partial^2\theta}{\partial t^2} - \frac{\partial}{\partial x}\left(EI\frac{\partial\theta}{\partial x}\right) - \frac{EA}{\mu_0}(1+\varepsilon)\left|B_0^r\right|\left|B^a\right|\sin(\alpha-\theta)$$

$$+\cos(\theta)\int_L^x \rho A \left[\int_0^x \left(\left(\frac{\partial\theta}{\partial t}\right)^2 \sin(\theta) - \frac{\partial^2\theta}{\partial t^2}\cos(\theta)\right)dx\right]dx$$

$$-\sin(\theta)\int_L^x \rho A \left[\int_0^x \left(\left(\frac{\partial\theta}{\partial t}\right)^2 \cos(\theta) + \frac{\partial^2\theta}{\partial t^2}\sin(\theta)\right)dx\right]dx = 0$$

(12a-b)

accompanied by the following boundary conditions at $x=0, L$

$$\varepsilon = 0 \text{ or } EA\varepsilon - \frac{A}{\mu_0}\left|B_0^r\right|\left|B^a\right|\cos(\alpha-\theta) = 0$$

$$\theta = 0 \text{ or } EI\frac{\partial\theta}{\partial x} = 0$$

(13a-b)

In view of Eq. (12a), $\varepsilon$ can be rewritten in terms of $\theta$. Substituting it into Eq. (12b) gives the following equation in terms of $\theta$

$$\rho I \frac{\partial^2\theta}{\partial t^2} - \frac{\partial}{\partial x}\left(EI\frac{\partial\theta}{\partial x}\right) - \frac{A}{\mu_0}\left|B_0^r\right|\left|B^a\right|\sin(\alpha-\theta) - \frac{A}{2E(\mu_0)^2}\left(\left|B_0^r\right|\left|B^a\right|\right)^2 \sin(2\alpha-2\theta)$$

$$+\cos(\theta)\int_L^x \rho A \left[\int_0^x \left(\left(\frac{\partial\theta}{\partial t}\right)^2 \sin(\theta) - \frac{\partial^2\theta}{\partial t^2}\cos(\theta)\right)dx\right]dx$$

$$-\sin(\theta)\int_L^x \rho A \left[\int_0^x \left(\left(\frac{\partial\theta}{\partial t}\right)^2 \cos(\theta) + \frac{\partial^2\theta}{\partial t^2}\sin(\theta)\right)dx\right]dx = 0$$

(14)

Calculating $\theta$ from the above equation and inserting it into Eq. (12a) lead to determining of $\varepsilon$. It is interesting to mention that the above equation reduces to that was obtained by Farokhi and Ghayesh [13] when $|B^a|$ is set to zero as well as constant cross section and Young's Modules for the beam are considered. Additionally, neglecting the time-dependent terms in Eq. (12a) or Eq. (14), the equations developed by Chen et al. [12] can be recovered. Eventually, dropping the time-dependent terms in Eq. (12b), considering constant material properties and geometrical parameters, and neglecting the centerline strain ($\varepsilon = 0$), the governing equation proposed by Wang et al. [8] can be attained.

It would be more convenient to rewrite Eqs. (12a), (14), and (10a-b) in a dimensionless form; to that end, first the following non-dimensional parameters for the beam with a uniform cross section are introduced

$$\zeta = \frac{x}{L}, \ \xi = \frac{u}{L}, \eta = \frac{v}{L}, \ \tau = \sqrt{\frac{EI}{\rho A L^4}}t, \ S = \frac{AL^2}{I}, \ P = \frac{AL^2 |B_0^r||B^a|}{EI \mu_0}, \ \delta x = \frac{\Delta x}{L}, \ \delta y = \frac{\Delta y}{L} \quad (15)$$

Substituting the above parameters into Eqs. (12a) and (14), the nonlinear formulation of the hard magnetic soft beam can be rewritten in dimensionless form:

$$\varepsilon - \frac{P}{S}\cos(\alpha - \theta) = 0$$

$$\frac{1}{S}\frac{\partial^2 \theta}{\partial \tau^2} - \frac{\partial^2 \theta}{\partial \zeta^2} - P\sin(\alpha - \theta) - \frac{P^2}{2S}\sin(2\alpha - 2\theta)$$

$$+ \cos(\theta)\int_1^\zeta \left[\int_0^\zeta \left(\left(\frac{\partial \theta}{\partial \tau}\right)^2 \sin(\theta) - \frac{\partial^2 \theta}{\partial \tau^2}\cos(\theta)\right)d\zeta\right]d\zeta \quad (16\text{a-b})$$

$$- \sin(\theta)\int_1^\zeta \left[\int_0^\zeta \left(\left(\frac{\partial \theta}{\partial \tau}\right)^2 \cos(\theta) + \frac{\partial^2 \theta}{\partial \tau^2}\sin(\theta)\right)d\zeta\right]d\zeta = 0$$

In addition, inserting Eqs. (15) and (16a) into Eq. (10a-b) yields the following non-dimensional displacements in terms of $\theta$:

$$\xi = \int_0^\zeta \left[\left(1 + \frac{P}{S}\cos(\alpha - \theta)\right)\cos(\theta) - 1\right]d\zeta$$

$$\eta = \int_0^\zeta \left(1 + \frac{P}{S}\cos(\alpha - \theta)\right)\sin(\theta)\, d\zeta \quad (17\text{a-b})$$

In order to analyze the nonlinear mechanics of hard magnetic soft beams, similar to the previous investigations, the quasi-static motion of the system is considered. To that end, the time-dependent terms in Eq. (16a-b) are dropped.

$$\varepsilon - \frac{P}{S}\cos(\alpha - \theta) = 0$$

$$\frac{d^2\theta}{d\zeta^2} + P\sin(\alpha - \theta) + \frac{P^2}{2S}\sin(2\alpha - 2\theta) = 0 \quad (18\text{a-b})$$

In order to obtain the deformed configuration of the beam, it is only enough to solve Eq. (16b). Since the shooting method is a powerful numerical technique to handle the boundary value problems, it is employed to determine the solutions of Eq. (18b). At the first step, the second-order ordinary differential equation given in Eq. (18b) must be converted into two first-order ordinary differential equations. Considering the following set of variables

$$\beta_1 = \theta, \ \beta_2 = \frac{d\theta}{d\zeta} \quad (19\text{a-b})$$

According to the above set of variables, Eq. (16b) can be converted into two first-order ordinary differential equation as follows

$$\frac{d\beta_1}{d\zeta} = \beta_2, \quad \frac{d\beta_2}{d\zeta} = P\sin(\beta_1 - \alpha) + \frac{P^2}{2S}\sin(2\beta_1 - 2\alpha) \qquad (20\text{a-b})$$

along with the following boundary conditions for a cantilever

$$\beta_1(0) = 0, \quad \beta_2(1) = 0 \qquad (21\text{a-b})$$

Now, the nonlinear shooting scheme for two-point boundary value problems [14] in conjunction with the fourth-order Runge-Kutta finite difference method and the Newton-Raphson method is applied on Eq. (20a-b) to determine the new variables introduced in Eq. (19a-b). Afterwards, the centerline strain ($\varepsilon$) can be directly calculated according to Eq. (18b). Additionally, in order to evaluate the dimensionless longitudinal and lateral displacements, the integrals in Eq. (17a-b) are numerically calculated via the adaptive quadrature technique [14].

## 3. Comparative studies

In this section, the results of experiments and finite element simulations of hard magnetic soft cantilevers under uniform perpendicular and antiparallel magnetic fields proposed by Zhao et al. [7] in addition to the analytical solution-based theoretical results of hard magnetic soft cantilevers under uniform magnetic field with arbitrary angles presented by Wang et al. [8] are employed to verify the solution strategy described in the previous section.

Figure 2 illustrates the free-end rotation angle ($\theta_L$) and the dimensionless free-end transverse location ($\delta y$) of hard magnetic soft cantilevers with different aspect ratios ($L/h$) under a uniform perpendicular magnetic field as a function of the normalized magnetic flux density ($P^* = |B_0^r||B^a|/G\mu_0 = 3P/S$). The plots in the figure indicate that the numerical solution-based theoretical results of the present work are in relatively good agreement with those reported by Zhao et al. [7] based on the experiments and finite element simulations. Besides, it can be inferred when the aspect ratio is increased, the differences lessen.

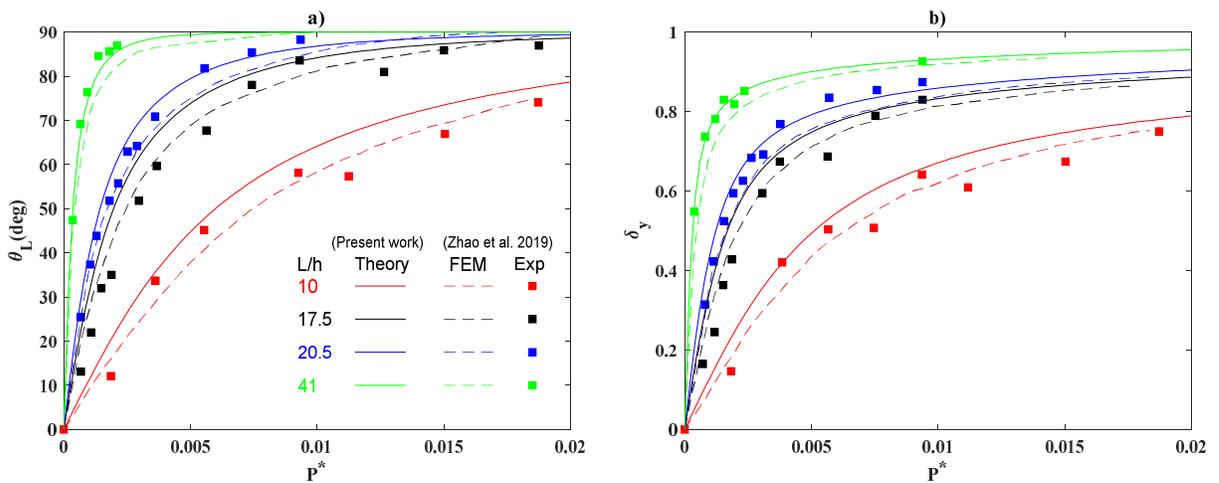

Figure 2: Comparative studies for a) the free-end rotation angle and b) the dimensionless free-end transverse location against the normalized magnetic flux density for the hard magnetic soft cantilevers with different aspect ratios exposed to a uniform perpendicular magnetic field ($\alpha = \pi/2$).

The geometric parameters of free-end of deformed configuration for a hard magnetic soft cantilever with *L/h=20.5* under a uniform antiparallel magnetic field against the normalized magnetic flux density are depicted in Figure 3. It can be seen that the results of the current study are in excellent agreement with those obtained by Wang et al. [8] and also they are in good agreement with those of Zhao et al. [7] based on the experiments and finite element simulations. It can be seen that both the theoretical models predict the pitchfork bifurcation for the beam under the uniform antiparallel magnetic field, but the finite element simulation conducted by Zhao et al. [7] doesn't predict such instability. It is probably due to the fact that employing the slightly inclined actuating magnetic field to make up for the absence of geometric imperfection or perturbation yields the inability of finite element simulation to capture the pitchfork bifurcation [8].

The plots in Figure 4a-b indicate that the results of the present work for $\theta_L$ and $\delta y$ of a hard magnetic soft cantilever under different *P* and $\alpha$ agree very well with those of Wang et al. [8] on the basis of the analytical solution-based theoretical model. Moreover, the depicted outcomes in Figure 4c demonstrate that the trajectories and workspace predicted by the current study are in very good agreement with those reported by Wang et al. [8]. It should be stressed that Wang et al. [8] stated that the aforementioned trajectories and associated workspace were obtained when $0 \leq \theta_L \leq \alpha$. But, the corresponding results based on this study are achieved when $0 \leq \theta_L \leq 0.999\alpha$. Besides, it should be pointed out that the results in Figure 4c are obtained by setting the centerline strain to zero.

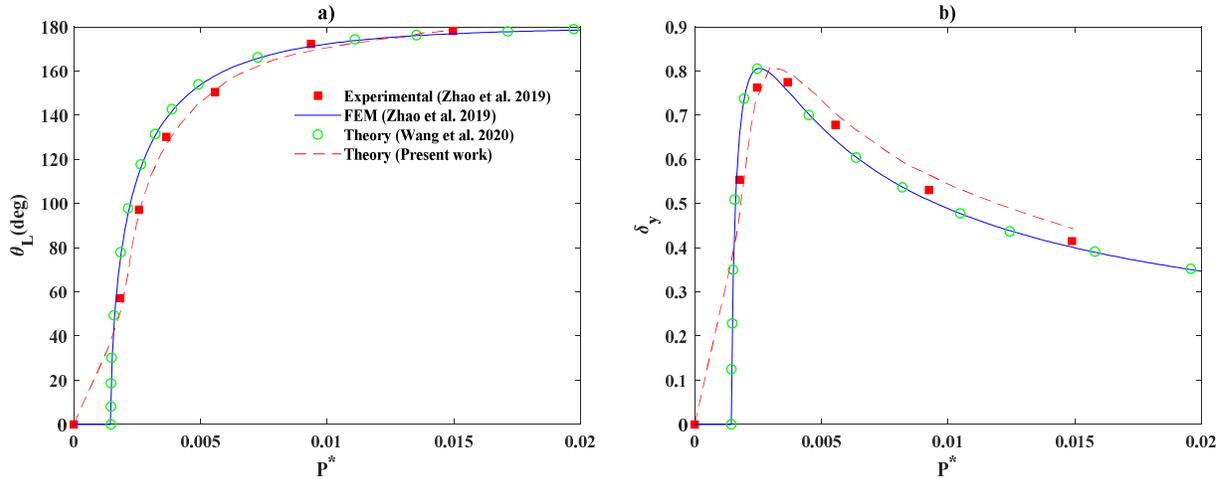

Figure 3: Comparative studies for a) the free-end rotation angle and b) the dimensionless free-end transverse location against the normalized magnetic flux density for a hard magnetic soft cantilever under a uniform antiparallel magnetic field ($\alpha = \pi$).

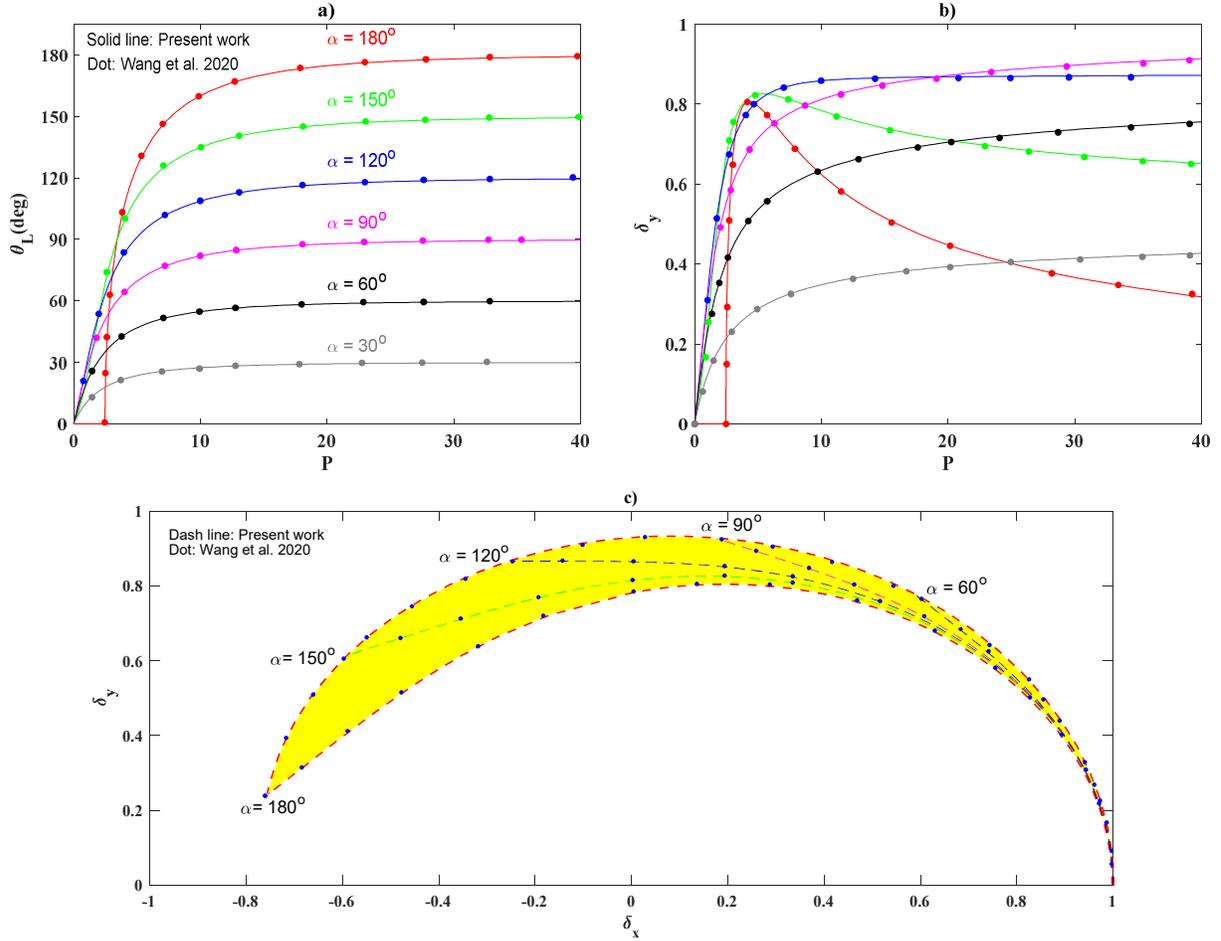

Figure 4: Comparative studies for a) the free-end rotation angle and b) the dimensionless free-end transverse location against the dimensionless parameter P for a hard magnetic soft cantilever subjected to a uniform magnetic field with different angles; c) the trajectories of free-end locations for different values of $\alpha$ as well as the work space (yellow region) of free-end locations of a hard magnetic soft cantilever when the magnitude and angle of the applied magnetic field vary.

## 4. Perturbed pitchfork bifurcation and new trajectories and workspace

In the previous studies [8, 10, 12], it was theoretically discussed when a hard magnetic soft cantilever is exposed to a magnitude field with an antiparallel angle ($\alpha = 180°$), the system undergoes a pitchfork bifurcation. However, for the other cases ($\alpha \neq 180°$), no bifurcation was predicted by the aforementioned references. Consider $\alpha = 180°$ and the system behaves based on the upper stable branch of pitchfork bifurcation (see Figure 4 for $\alpha = 180°$), if the angle $\alpha$ is gradually varied, the free-end of system slowly moves in the yellow region depicted in Figure 4c. But, when the system behaves based on the lower stable branch of pitchfork bifurcation, any small variation in the angle $\alpha$ leads to a great and abrupt movement to return the system to a stable position. In view of this scenario, the current section is devoted to finding the possible bifurcation and trajectories around the aforementioned lower branch for $\alpha \neq 180°$.

In order to obtain a better understanding of the nonlinear mechanics of beam, at the first step, the nonlinear responses of the system under an antiparallel angle ($\alpha = 180°$) is examined by considering possible bifurcations. To that end, the free-end rotation angle ($\theta_L$) along with the free-end transverse location ($\delta y$) of a hard magnetic soft cantilever with $L/h=20.5$ under a constant magnetic field with an antiparallel angle is illustrated in Figure 5a-b. The plotted results show that the system may undergo pitchfork bifurcation when $P^* = 0.00147$ or $0.01317$. As discussed in the beginning of this section, the first pitchfork bifurcation was predicted in the preceding investigations. Generally, the system may undergo more pitchfork bifurcations when $P^*$ is increased. In the case of $\alpha = 180°$, Eq. (18b) can be rewritten

$$\frac{d^2\theta}{d\zeta^2} + P\sin(\theta) - \frac{P^2}{2S}\sin(2\theta) = 0 \tag{22}$$

It is obvious that $\theta = 0$ is a trivial solution of Eq. (22) for each arbitrary value of dimensionless parameter $P$. Using the Taylor series of sine function, Eq. (22) can be rewritten as

$$\frac{d^2\theta}{d\zeta^2} + P\left(\theta - \frac{\theta^3}{3!} + \frac{\theta^5}{5!} - \cdots\right) + \frac{P^2}{2S}\left(2\theta - \frac{(2\theta)^3}{3!} + \frac{(2\theta)^5}{5!} - \cdots\right) = 0 \tag{23}$$

Eq. (23) can be linearized around $\theta = 0$ for predicting pitchfork bifurcation

$$\frac{d^2\theta}{d\zeta^2} + \left(P - \frac{P^2}{S}\right)\theta = 0 \tag{24}$$

The solution of Eq. (24) can be obtained by

$$\theta = A\sin\left(\sqrt{P - \frac{P^2}{S}}\zeta\right) + B\cos\left(\sqrt{P - \frac{P^2}{S}}\zeta\right) \tag{25}$$

For a cantilever with the boundary conditions listed in Eq. (21a-b), the pitchfork bifurcation occurs when

$$P^* = \frac{3}{2}\left(1 - \sqrt{1 - \frac{(n\pi)^2}{S}}\right), \quad n=1,3,5,\ldots \tag{26}$$

Based on Eq. (26), the first and second pitchfork bifurcations take place when $P^* = 0.00147$ and $0.0133$, respectively. They are in very good agreement with those obtained by the numerical method.

The corresponding deformed configurations of the system for the non-zero branches (B2-B5) are shown in Figure 5c-d. From the plots, it can be seen that the deformed configurations of the first pitchfork bifurcation are totally different from those of the second one. The possibility of the occurrence of the second pitchfork bifurcation will be discussed later by using the potential energy function.

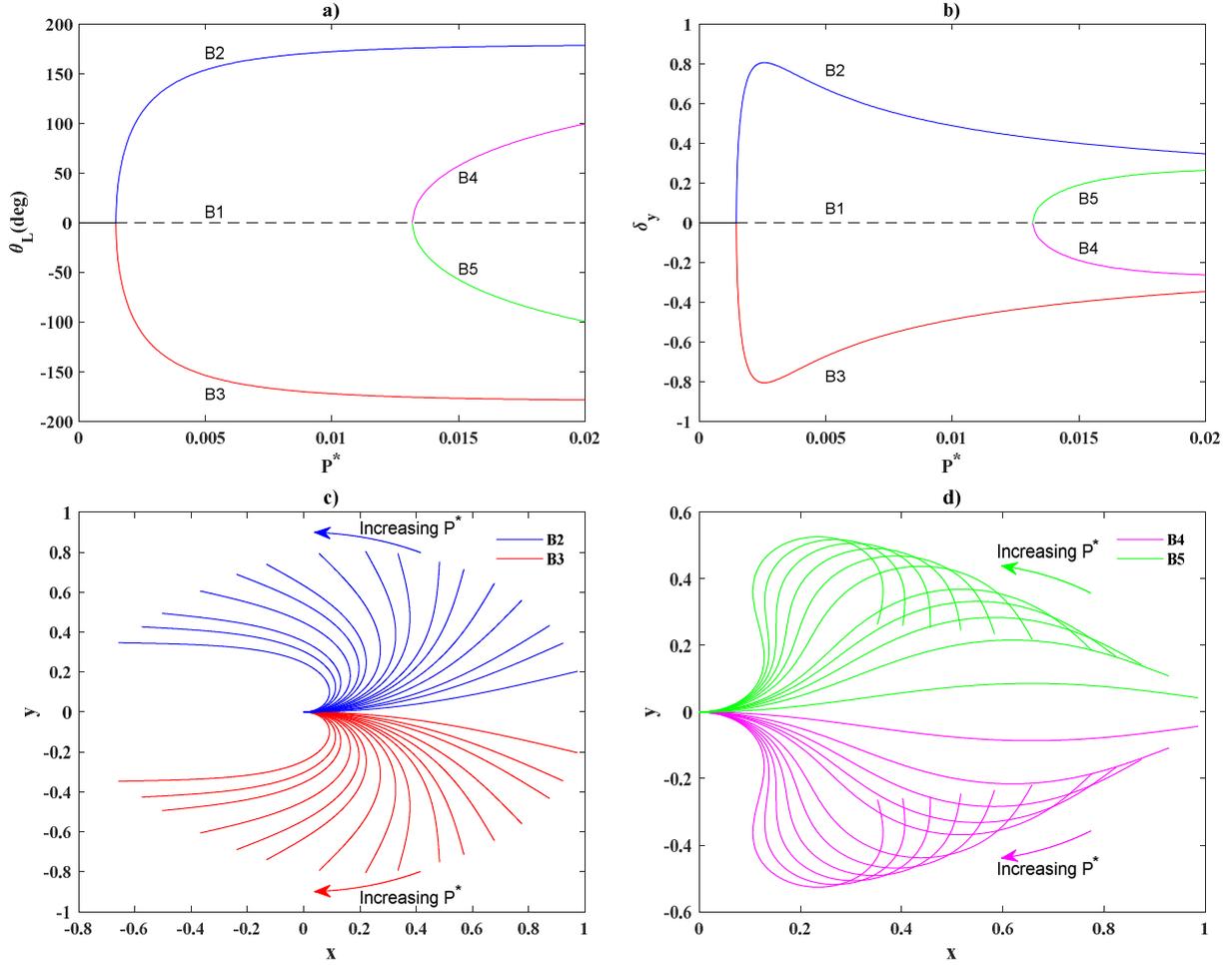

Figure 5: a) The free-end rotation angle and b) the dimensionless free-end transverse location as a function of the normalized magnetic flux density for a hard magnetic soft cantilever subjected to a constant antiparallel magnetic field ($\alpha = 180°$); the deformed configurations of the beam when the normalized magnetic flux density increases for c) the second and third stable branches, d) the fourth and fifth stable branches.

Based on the Taylor series of sine and cosine functions, Eq. (18b) can be rewritten as

$$\frac{d^2\theta}{d\zeta^2} + P\sin(\alpha)\left(-\frac{\theta^2}{2!} + \frac{\theta^4}{4!} - ...\right) - P\cos(\alpha)\left(\theta - \frac{\theta^3}{3!} + \frac{\theta^5}{5!} - ...\right)$$
$$+ \frac{P^2}{2S}\sin(\alpha)\left(-\frac{(2\theta)^2}{2!} + \frac{(2\theta)^4}{4!} - ...\right) - \frac{P^2}{2S}\cos(\alpha)\left(2\theta - \frac{(2\theta)^3}{3!} + \frac{(2\theta)^5}{5!} - ...\right) = -\left(P + \frac{P^2}{2S}\right)\sin(\alpha) \quad (27)$$

Due to the presence of the non-zero term in the right-hand side of Eq. (27) for $\alpha \neq 0, 180°$, it is expected that the system experiences perturbed pitchfork bifurcation rather than pitchfork bifurcation when the angle $\alpha$ is not parallel or antiparallel to *x*-axis. In order to investigate the nonlinear mechanics of system for $\alpha \neq 0, 180°$, the free-end rotation angle and dimensionless free-end transverse angle of a hard magnetic soft beam under a constant magnetic field with $\alpha = 170°$ are illustrated in Figure 6a-b, respectively.

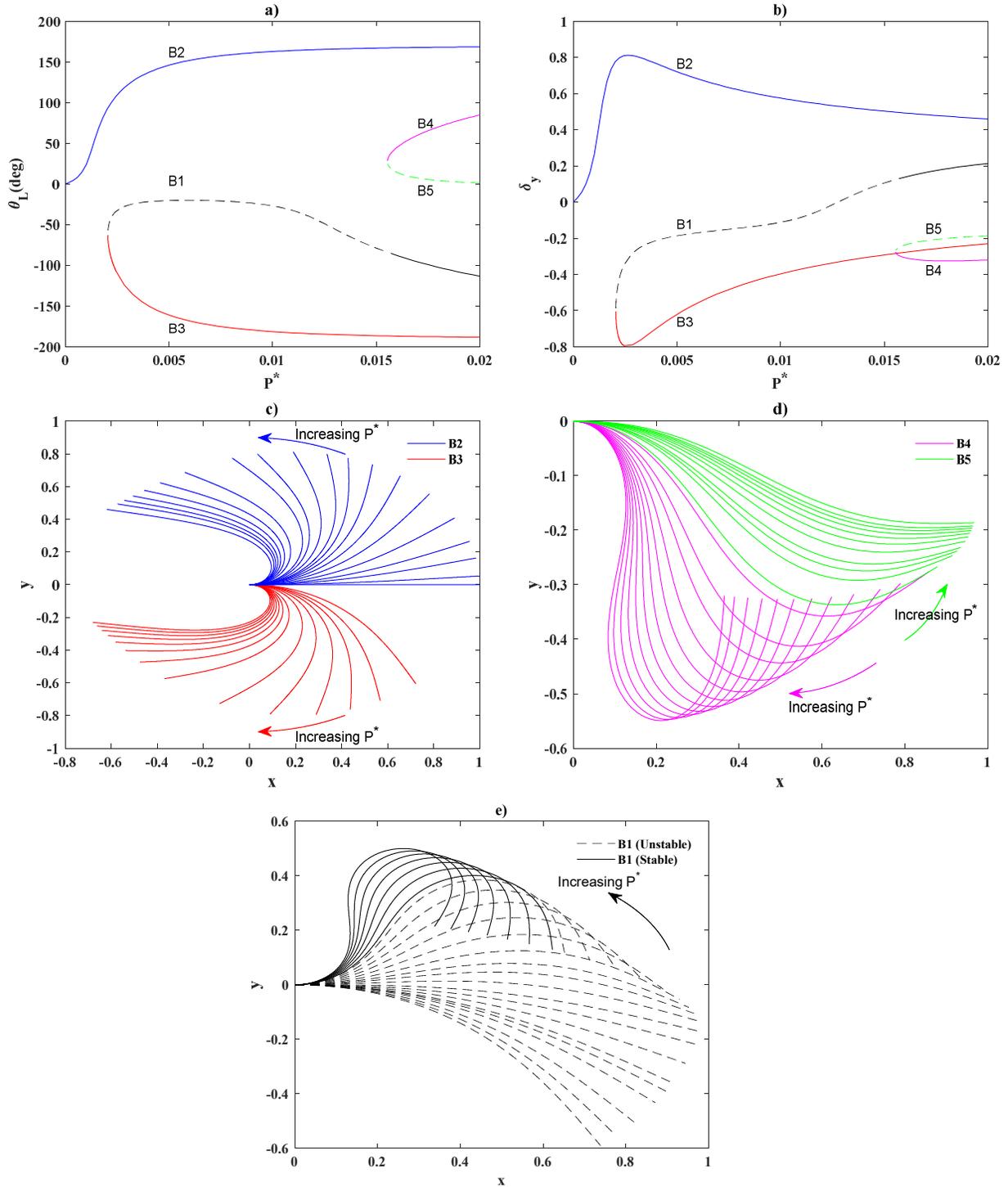

Figure 6: a) The free-end rotation angle and b) the dimensionless free-end transverse location against the normalized magnetic flux density for a hard magnetic soft cantilever under a constant magnetic field with $\alpha = 170°$; the deformed configurations of the beam when the normalized magnetic flux density increases for c) the second and third stable branches, d) the fourth stable branch and fifth unstable branch, and e) the first branch.

By comparing the plots in Figure 5a-b with those in Figure 6a-b, it can be inferred that the general behavior of the system with $\alpha = 170°$ is different from that with $\alpha = 180°$. From the plots in Figure 6a-b, it can be seen that the system with $\alpha = 170°$ undergoes the perturbed pitchfork bifurcations at $P^* = 0.00205$ or $0.01555$. In the first perturbed pitchfork bifurcation, the first branch ($B_1$) is unstable, but it becomes stable when the second perturbed pitchfork bifurcation takes place. The deformed configurations of the system with $\alpha = 170°$ as a function of the normalized magnetic flux density are plotted in Figure 6c-e.

In order to analyze the possibility of occurrence of different stable branches, the dimensionless potential energy $U^* = (U_s + U_m)L/EI$ of the system versus the normalized magnetic flux density are illustrated in Figure 7 for $\alpha = 180°$ and $170°$. In the case of $\alpha = 180°$, it can be inferred when $P^* > 0.01317$ i.e., after occurring the first pitchfork bifurcation, the possibility of occurrence of the second or third branch is much more than the fourth or fifth branch due to their low level of potential energy. In other words, it can be stated that both the second and third branches are global stable solutions when $P^* > 0.00147$. In the case of $\alpha = 170°$, the plotted dimensionless energy functions show that the second branch is the global stable solution. In light of the dimensionless potential energy function of the third branch and its deformed configuration (see Figure 6c), after occurring the first perturbed pitchfork bifurcation, this branch can be considered as a local stable solution with the high possibility of occurrence when the system is subjected to appropriate initial conditions. Moreover, the other stable solutions including the first and fourth branches which appear after taking place the second perturbed pitchfork bifurcation, can be regarded as a local stable solution with a low possibility of occurrence owing to the high level of potential energy with respect to the other stable solutions. Consequently, it can be concluded that investigation on the behavior of the third branch can be very handy for the analysis and design of hard magnetic soft beams.

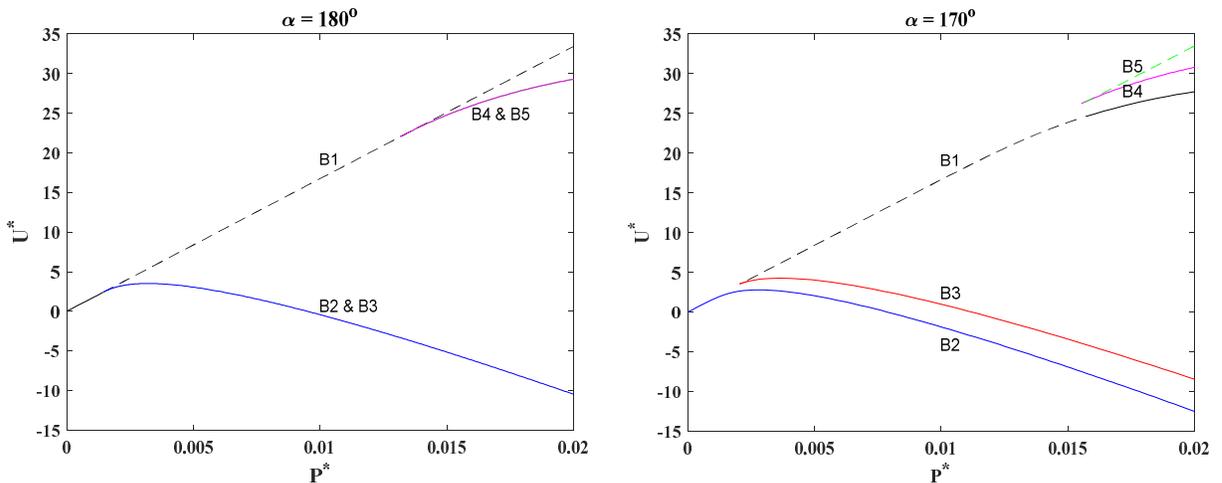

Figure 7: The dimensionless potential energy of the beam against the normalized magnetic flux density for different branches.

Figure 8 is constructed to give better insights to the first perturbed pitchfork bifurcation and the starting points of the third branch as a local stable solution with a high possibility of occurrence in respect of the other local stable solutions. The plots in Figure 8a-b display the locus of starting points of the first bifurcation in $\theta_L$-$P^*$ plane and $\delta y$-$P^*$ plane, respectively. The plot in Figure 8a indicates when the angle $\alpha$ is decreased, the starting point of first bifurcation takes place at a higher normalized magnetic flux density ($P^*$) as well as the non-positive value of free-end rotation angle ($\theta_L$) of the starting point reduces. Additionally, based on the plot in Figure 8b, the free-end dimensionless transverse location ($\delta y$) of the beam has an initial descending trend with respect to reducing the angle $\alpha$ which is followed by an ascending trend. From the results plotted in the figure, two important conclusions can be drawn. The first one is that in the beginning of the first bifurcation, the beam undergoes larger deformation when the angle $\alpha$ is decreases. The second one is that the beam has only one stable solution for a greater interval of $P^*$ when the angle $\alpha$ is lessened.

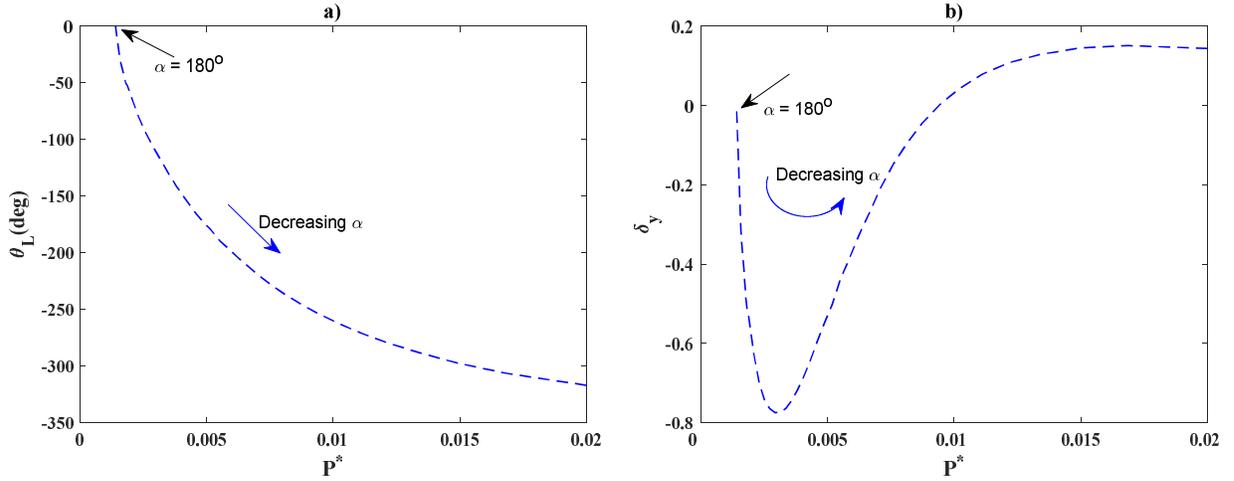

Figure 8: The locus of starting points of first bifurcation or third branch in $\theta_L$-$P^*$ and $\delta_y$-$P^*$ planes when the angle $\alpha$ varies.

In the case of the second branch, when the normalized magnetic flux density ($P^*$) is increased, the free end rotation angle ($\theta_L$) rises monotonically and finally tends to $\alpha$ (see Figure 4a). It should be mentioned that Wang et al. [8] expressed that based on their analytical solution, the free end rotation angle approaches to $\alpha$, but the results of present work show that the free end rotation angle tends to $\alpha$ and does not touch $\alpha$. When $\alpha \leq 90°$, the dimensionless free-end transverse location ($\delta y$) increases monotonically and then becomes saturated as the normalized magnetic flux density is increased (see Figure 4b). Additionally, when $\alpha > 90°$, the dimensionless free-end transverse location has an ascending trend with respect to increasing the normalized magnetic flux density which is followed by a descending trend (see Figure 4b). The rate of decrease in the descending trend rises when the angle $\alpha$ is increased. Figure 9 shows the free-end rotation angle ($\theta_L$) as well as the dimensionless free-end transverse location ($\delta y$) of the third branch for the hard magnetic soft cantilever exposed to a uniform magnetic field when the normalized magnetic flux density ($P^*$) and the angle $\alpha$ vary. According to the plots in Figure 9a, in the case of the third branch, as the normalized magnetic flux density is increased, the free-end rotation angle decreases monotonically and lastly tends to $-2\pi + \alpha$. The plots in Figure 9b show that for $\alpha > 150°$, when the normalized magnetic flux density is increased, the dimensionless free-end transverse location initially lessens and then rises after reaching its minimum. Besides, when $\alpha \leq 150°$, the dimensionless free-end transverse location rises monotonically and then becomes saturated as the normalized magnetic flux density is increased.

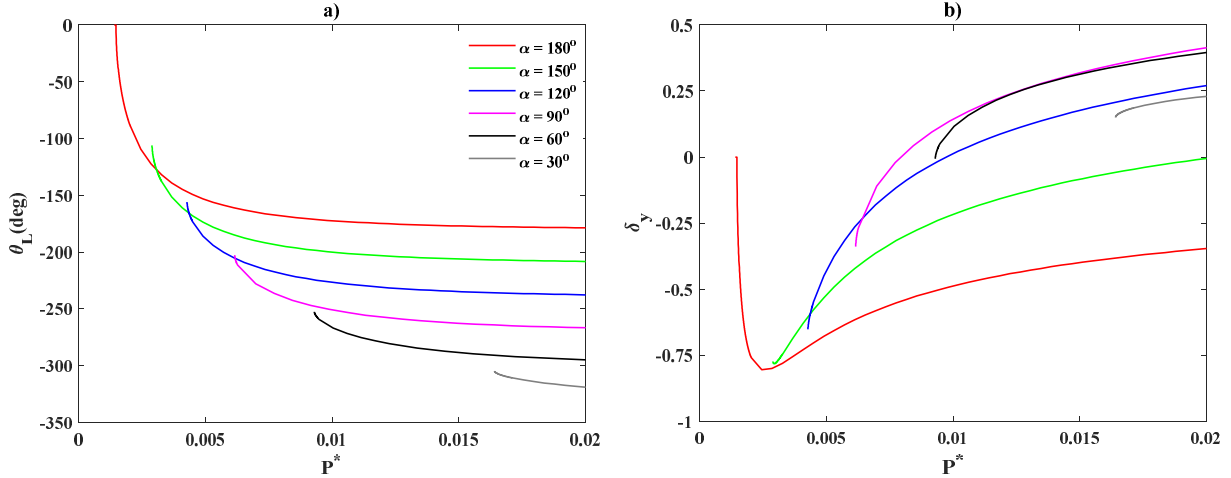

Figure 9: a) The free-end rotation angle and b) the dimensionless free-end transverse location versus the normalized magnetic flux density for the third branch of a hard magnetic soft cantilever exposed to a uniform magnetic field with various angles.

Figure 10 shows the third branch-based trajectories (the dash lines) of free-end locations of the hard magnetic soft cantilever for different angles $\alpha$ when the normalized magnetic flux density ($P^*$) is varied. Additionally, the third branch-based workspace (the yellow region) of the system is depicted by varying the normalized magnetic flux density and the angle $\alpha$. It should be pointed out that for each prescribed angle $30^0 \leq \alpha \leq 180^0$, the minimum normalized magnetic flux density of the one at which the first bifurcation takes place (see Figure 8) and the maximum normalized magnetic flux density of the one at which $\theta_L$ equals $0.999(-2\pi + \alpha)$. The lower and upper dash-dot curves in Figure 10 correspond to the minimum and maximum normalized magnetic flux densities, respectively. Consider the upper dash-line curve related to the maximum normalized magnetic flux density, when the angle $\alpha$ is reduced, the beam undergoes larger deformation, but a reverse scenario happens for the second branch (see Figure 4c). Based on the lower dash-line curve related to the minimum normalized magnetic flux density, the beam under lower angle $\alpha$ experiences larger deformation. It should be noticed that in the case of the second branch (see Figure 4a-c), the point $(\delta x, \delta y) = (1,0)$ corresponds to the minimum normalized magnetic flux density for each arbitrary angle $\alpha$.

There exists a major difference between the performances of the system in the third branch-based workspace plotted in Figure 10 and the second branch-based workspace plotted in Figure 4c. In the second-branch-based workspace, when $P^*$ or $\alpha$ is slightly varied, the free-end of beam moves in the yellow region depicted in Figure 4c, but for the third branch-based workspace, a slight change of $P^*$ or $\alpha$ may result in a predictable but abrupt and large movement towards the second branch-based workspace. It is due to the fact that, for each prescribed angle $\alpha$, when $P^*$ is smaller than that at which the first bifurcation occurs, the second-branch is only stable response of system. In this scenario, there is no solution in the third branch-based workspace and the beam finds its stable free-end location in the second branch-based workspace. This feature of the third branch-based workspace can be regarded as an advantage to develop mechanisms that need to have restrictive boundaries.

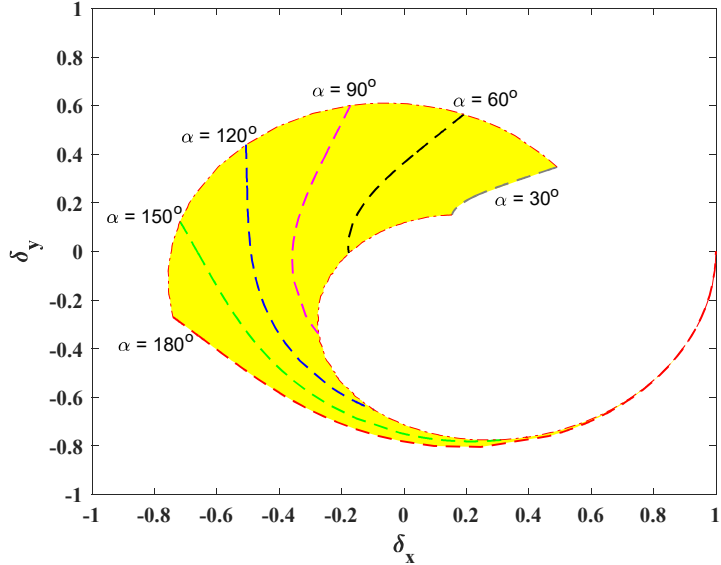

Figure 10: The third branch-based trajectories of free-end locations of a hard magnetic soft cantilever for different angles $\alpha$ in addition to the third branch-based workspace (yellow region) of the free-end of the cantilever when the normalized magnetic flux density $P^*$ and angle $\alpha$ are varied.

## 5. Conclusions

In the framework of the current investigation, the exact geometric nonlinear mechanics of hard magnetic soft cantilevers exposed to an actuating uniform magnetic field is examined theoretically and numerically via taking into account bifurcation. It is analytically indicated that the system under antiparallel magnetic field undergoes pitchfork bifurcation at different normalized magnetic flux densities. Additionally, the analysis shows that the system undergoes perturbed pitchfork bifurcations when the magnetic field is not parallel or antiparallel to the undeformed configuration of the cantilever. The potential energy functions of the system show that the local stable response due to the first perturbed pitchfork bifurcation has a high possibility of occurrence with respect to the other local stable solutions. The findings reveal that when the angle of the applied magnetic field is increased, the first bifurcation occurs at a lower normalized magnetic flux density. Moreover, at the beginning of the first perturbed pitchfork bifurcation, the beam undergoes larger deformation when the angle of the applied magnetic field is reduced. The new trajectories of the free-end location of hard magnetic soft cantilevers due to the first perturbed pitchfork bifurcation are presented and compared with those obtained in the previous study [8] based on the global stable response. The new trajectories propose a different workspace for the free-end location of the hard magnetic soft cantilever which has a main difference from that presented in the preceding investigation [8]. The difference is that in the new workspace, a gradual variation may lead to a predictable but large and abrupt movement. This characteristic can be employed to produce mechanisms with restrictive boundaries.

## References

[1] Y. Kim, H. Yuk, R. Zhao, S. A. Chester, and X. Zhao, "Printing ferromagnetic domains for untethered fast-transforming soft materials," *Nature,* vol. 558, no. 7709, pp. 274-279, 2018.

[2] W. Hu, G. Z. Lum, M. Mastrangeli, and M. Sitti, "Small-scale soft-bodied robot with multimodal locomotion," *Nature,* vol. 554, no. 7690, pp. 81-85, 2018.

[3] Y. Kim, G. A. Parada, S. Liu, and X. Zhao, "Ferromagnetic soft continuum robots," *Science Robotics,* vol. 4, no. 33, 2019.

[4] R. Zhang, S. Wu, Q. Ze, and R. J. J. o. A. M. Zhao, "Micromechanics Study on Actuation Efficiency of Hard-Magnetic Soft Active Materials," vol. 87, no. 9, 2020.

[5] N. Bira, P. Dhagat, and J. R. Davidson, "A Review of Magnetic Elastomers and Their Role in Soft Robotics," *Frontiers in Robotics and AI,* vol. 7, p. 146, 2020.

[6] S. Wu, C. M. Hamel, Q. Ze, F. Yang, H. J. Qi, and R. Zhao, "Evolutionary Algorithm-Guided Voxel-Encoding Printing of Functional Hard-Magnetic Soft Active Materials," *Advanced Intelligent Systems,* p. 2000060, 2020.

[7] R. Zhao, Y. Kim, S. A. Chester, P. Sharma, and X. Zhao, "Mechanics of hard-magnetic soft materials," *Journal of the Mechanics and Physics of Solids,* vol. 124, pp. 244-263, 2019.

[8] L. Wang, Y. Kim, C. F. Guo, and X. Zhao, "Hard-magnetic elastica," *Journal of the Mechanics and Physics of Solids,* vol. 142, p. 104045, 2020.

[9] S. Wu *et al.*, "Symmetry-Breaking Actuation Mechanism for Soft Robotics and Active Metamaterials," *ACS applied materials & interfaces,* vol. 11, no. 44, pp. 41649-41658, 2019.

[10] W. Chen and L. Wang, "Theoretical modeling and exact solution for extreme bending deformation of hard-magnetic soft beams," *Journal of Applied Mechanics,* vol. 87, no. 4, 2020.

[11] W. Chen, Z. Yan, and L. Wang, "Complex transformations of hard-magnetic soft beams by designing residual magnetic flux density," *Soft Matter,* 2020.

[12] W. Chen, Z. Yan, and L. Wang, "On mechanics of functionally graded hard-magnetic soft beams," *International Journal of Engineering Science,* vol. 157, p. 103391, 2020.

[13] H. Farokhi and M. H. Ghayesh, "Extremely large-amplitude dynamics of cantilevers under coupled base excitation," *European Journal of Mechanics-A/Solids,* vol. 81, p. 103953, 2020.

[14] R. L. Burden, D. J. Faires, and A. M. Burden, *Numerical analysis*, Tenth ed. Cengage, 2016.